# A setup for the study of surface processes under external homogeneous electric fields

*W. Steurer, S. Surnev, F. Hanauer, K. Ansperger, F.P. Netzer*

*Surface and Interface Physics, Institute of Physics, Karl-Franzens University Graz,*

*A-8010 Graz, Austria*

**Abstract**

A setup for studying the influence of external electric fields on dynamic surface processes is described. Spatially-extended homogeneous electric fields are realized by applying a DC voltage in between a planar electrode and a metallic substrate, which are separated by a narrow vacuum gap. The functional capability of the device is demonstrated by recording the field-emission characteristic as a function of the applied field up to 2.3 V/nm, revealing proper Fowler-Nordheim tunneling behavior.

**1. Introduction**

In surface nanoscience, the influence of factors such as electric and magnetic fields have largely been neglected in the study of dynamic processes such as growth, reactivity, and freezing and melting. The influence of electric fields on surface reactions has traditionally been studied using field ion microscopy [1-3]. *Inhomogeneous* electric fields >10V/nm can be realized at the apex of a sharp tip but investigations are inherently limited to positive tip biases. With the advent of scanning tunneling microscopy and atomic force microscopy, studies have concentrated on understanding the influence of the inhomogeneous electric field underneath the probe tip on measured surface properties, such as diffusion rates. A multitude of tip-induced effects has been reported [4-14], but a quantitative evaluation of the results has proven to be difficult, mainly because of the ill-defined geometry of the tip. Recently, it was proposed theoretically that the structure of metal clusters deposited on non-polar ultrathin oxide films is strongly dependent on the work function of the underlying metal [15]. This opens up the possibility to orient the cluster growth by polarizing the support via an electric bias. In addition, a number of other field-induced effects have been predicted by theory [16-24], demonstrating the recent interest in this field and the necessity of experimental studies.

Here we describe a new setup with which large surface areas can be exposed to *homogeneous* external electric fields as presupposed in all theoretical predictions. We demonstrate the functional capability of the device by showing Fowler-Nordheim plots of field-emitted electrons as a function of the applied field up to 2.3 V/nm.

## 2. Instrument design

The assembly of the electric-field setup is shown in Fig. 1. It consists of a sample holder with electron-beam heating and a thermocouple for temperature reading, an electrode holder containing a miniature electrode from Litcon Ltd. attached to a manual linear motion manipulator, and a wobble stick for sample transfer. The whole setup is mounted inside an ultra-high vacuum (UHV) chamber. All components are electrically isolated from ground. The electrode holder and the sample holder are wired to the outside of the chamber via high-voltage UHV-feedthroughs. The miniature electrode shown in the inset of Fig. 1 is made out of quartz ($SiO_2$) with a gold stripe evaporated on top of it. A schematic drawing of the miniature electrode is presented in Fig. 2a. The gold electrode in the center is lowered by 800 nm with respect to the outer support legs. For reasons described later, the gold electrode is surrounded by a narrow groove of 0.3 mm depth. The gold stripe is 3x5 mm² in dimension. The gold electrode is electrically contacted to the electrode holder either by wire bonding or by using silver conducting epoxy. A detailed sketch of the electrode holder is shown in Fig. 2b. It consists of a piston suspended from a shaft which is mounted on the linear motion manipulator and Cu-Be springs which secure the exchangeable electrode holder plates. The latter contains the miniature electrode fastened by a fitting piece. A high-voltage power supply from Bertan Associates Inc. (model 205A-03R) with switchable polarity is used for applying a bias between the miniature electrode and the sample. In order to prevent the miniature electrode and the power supply from damages due to potential electrical breakdowns in the gap region at high field strengths, a current limiter (see wiring diagram displayed in Fig. 3; set to a maximum of 0.2 mA) is introduced into the electric circuit.

For an experiment, the sample is transferred from the UHV chamber's main manipulator to the sample holder. Thereafter, the manipulator is retracted and the electrode holder with the attached miniature electrode is lowered from above onto the sample. After landing on the sample, the suspended piston of the electrode holder assures that the miniature electrode stands force-free with both legs on the sample surface. This is important as it ensures a uniform gap in between the gold coating of the miniature electrode and the sample surface of 800 nm; the latter is given by the level difference between the outer support legs and the gold electrode in the center.

The instrument design is based on the following considerations. While the basic form – consisting of two plane-parallel electrodes – is naturally imposed by the goal of applying homogeneous electric fields, the projected field strength of 2 V/nm rules out a classical plate capacitor design in which the sample is sandwiched in between the electrodes. Convenience and safety considerations limit admissible voltages to below 10 kV. The voltage range is further restricted to below 2 kV by the expected field emission of 0.1–100 µA at the planned field strengths and the related unwanted effects. These range from anode heating, the knock-out of single atoms on the anode, to the point of electrical breakdown and anode destruction. The energy intake by electrons is proportional to the applied voltage but the destructive character of the electrons impinging on the anode increases greatly above 2 kV. The permissible voltage range imposes a gap between the electrodes of less than one micrometer. Considering that a typical sample in surface nanoscience is of the order of 1 mm thick, fields of the projected strength can only be achieved if the sample itself is metallic and acts as one of the electrodes. Furthermore, the narrow gap in between the electrodes requires high demands on the surface finish of both the anode and the cathode and, even more importantly, on their precise plane-parallel alignment. We achieve the latter by the particular design of the miniature electrodes whose support legs serve as "spacers". The lowered plane of the gold coating is manufactured using reactive ion etching technique, hereby ensuring a very high degree of parallelism and low surface roughness values <10 nm. Because the break-down strength of quartz is two orders of magnitude less than the electric field strength in the experiment (E~ 1-2 V/nm), the gold area is surrounded by 0.3 mm deep, narrow grooves which prolong the field lines in the quartz frame and push the maximum field strength along the field lines below the break-down value. In order to account for electrode deterioration effects (uncontrolled field breakdowns, heat-up and ablation of material due to high field-emission currents at local surface roughnesses, etc.) the instrument is designed in a way that miniature electrodes can be loaded into the UHV chamber and quickly exchanged.

3. Results

In order to demonstrate the functional capability of the device we have measured the field emission current as a function of the applied voltage on a freshly polished Ag(100) crystal surface with a maximum roughness height ≤30 nm. A Fowler-Nordheim (FN) plot [ln(I) vs. $V^{-1}$] up to a field strength of 2.3 V/nm is shown in Fig. 5. The solid red line is a fit to the experimental data (open circles) using the general-barrier FN equation [25]

$$I(F) = \left[\left[v\left(\frac{F}{F_\phi}\right) - \frac{2}{3}\phi\right]\right]^{-2} A a \phi^{-1} F^2 \exp\left\{-\frac{v\left(\frac{F}{F_\phi}\right) b \phi^{\frac{3}{2}}}{F}\right\} \quad (1)$$

where $A$ is the emission area, $\phi$ the local work function, $F$ the applied field strength, and $a = 1.541434 \times 10^{-6} \text{A eV V}^{-2}$ and $b = 6.83089 \text{ eV}^{-3/2} \text{V nm}^{-1}$ are the first and the second Fowler-Nordheim constants. In (1)

$$v\left(\frac{F}{F_\phi}\right) = 1 - \frac{F}{F_\phi} + \frac{1}{6}\frac{F}{F_\phi} \ln \frac{F}{F_\phi} \quad (2)$$

with $F_\phi = (0.6944617 \text{Vnm-1})\left(\frac{\phi}{\text{eV}}\right)^2$, the field needed to reduce to zero a Schottky-Nordheim barrier of unreduced height equal to the local work-function $\phi$. For the fitting procedure, $F$ is replaced by $\frac{U}{d}$, where $U$ is the applied voltage, and the gap $d$ between the electrodes is treated as a fit parameter. Hereby we take into account that on a real surface the predominant portion of the field-emitted electrons will originate from a restricted area with increased local surface roughness and/or smallest electrode separation, i.e., where the local field is strongest. The emission area is the only other fit parameter. With $\phi = 4.64\,eV$ for Ag(100) [26], excellent agreement with the experimental data ($\chi^2 = 0.04$, $R^2 = 0.998$) is obtained for an emission area of $A = 3 \pm 2 \times 10^6\,nm^2$ and a gap of d = 450 nm. These values indicate that the main emitters on the surface are particles, 200-350 nm in size, which locally reduce the gap separation. As the sample is transferred between the individual stages (sputtering, E-field, STM) several times per day, contamination of the single-crystal surface with small particles cannot be prevented completely. Especially the approach of the miniature electrode from above bears a certain risk that microparticles come off the linear motion mechanism and land on the sample surface. Ar+ sputtering, which is the only method to clean the Ag(100) crystal surface in UHV, is very inefficient for removing particles. E. g., hundreds of sputter cycles are required to completely remove a particle that is 300 nm in size. That the applied electric field is indeed very homogeneous over the whole electrode area is confirmed by our observation of electric-field induced effects in ultrathin oxide films on metal supports. For fields exceeding a critical threshold, large-scale reduction of the oxide film in the area covered by the electrode has been found, a result which will be presented in a forthcoming publication [27].

## 4. Conclusions

We have developed a new experimental setup dedicated to the study of the influence of homogeneous electric-fields on surface properties. By applying a bias between a flat, metallic sample and a custom-built miniature electrode with a gap distance of 800 nm we achieve field strengths up to 2.3 V/nm as evidenced by the measured field-emission characteristic. This opens up new possibilities in surface nanoscience with the prospects of tuning surface properties by electric fields and creating novel nanostructures. Furthermore, many of the predicted field-induced surface effects [16-24] can be verified experimentally with the described setup.

**Acknowledgement**

This work has been financed by the ERC Advanced Grant SEPON.

**Figure captions**

**Figure 1:** Photograph of the experimental setup taken during sample transfer from the main manipulator to the field sample-holder stage. The colored insert photo shows a miniature electrode from Litcon Ltd. prior to mounting (scale bar = 2 mm).

**Figure 2:** Schematic drawing of a miniature electrode (a) and assembly of the electrode holder (b).

**Figure 3:** Wiring diagram. The transistors together with the resistors act as a current limiter in case of a field breakdown.

**Figure 4:** Fowler-Nordheim field-emission characteristic. The red curve is a fit to the experimental data points (open circles) using Eq. 1. The top abscissa, indicating the applied field strength, is given for the nominal gap distance of 800 nm.

Figure 1:

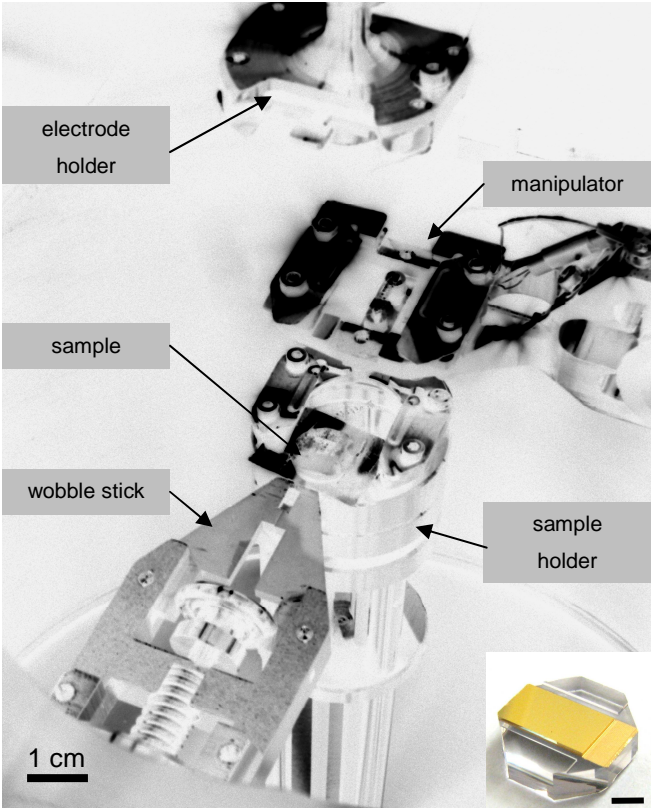

Figure 2:

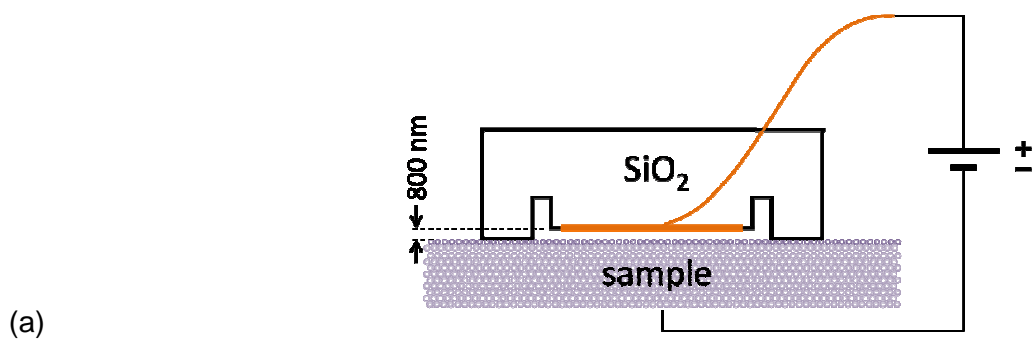

(a)

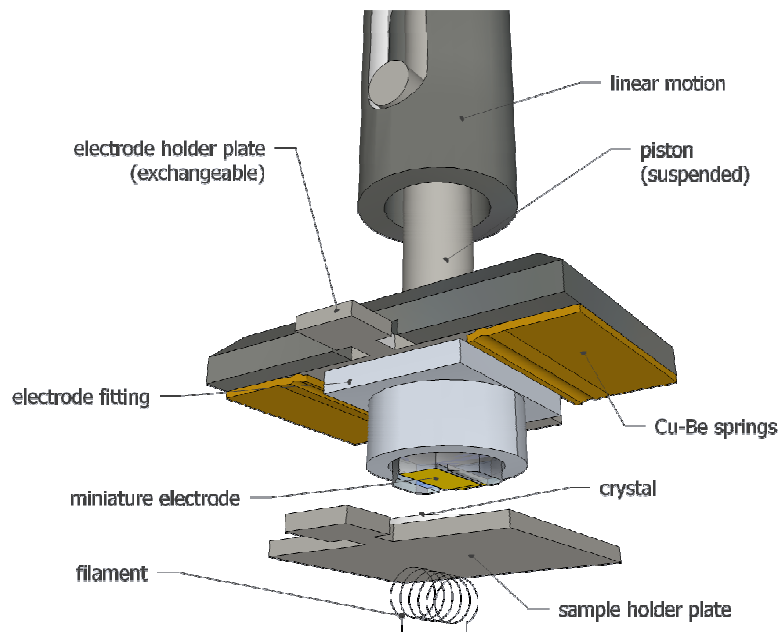

(b)

Figure 3:

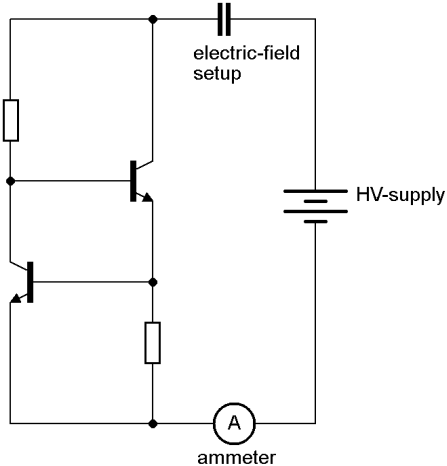

Figure 4:

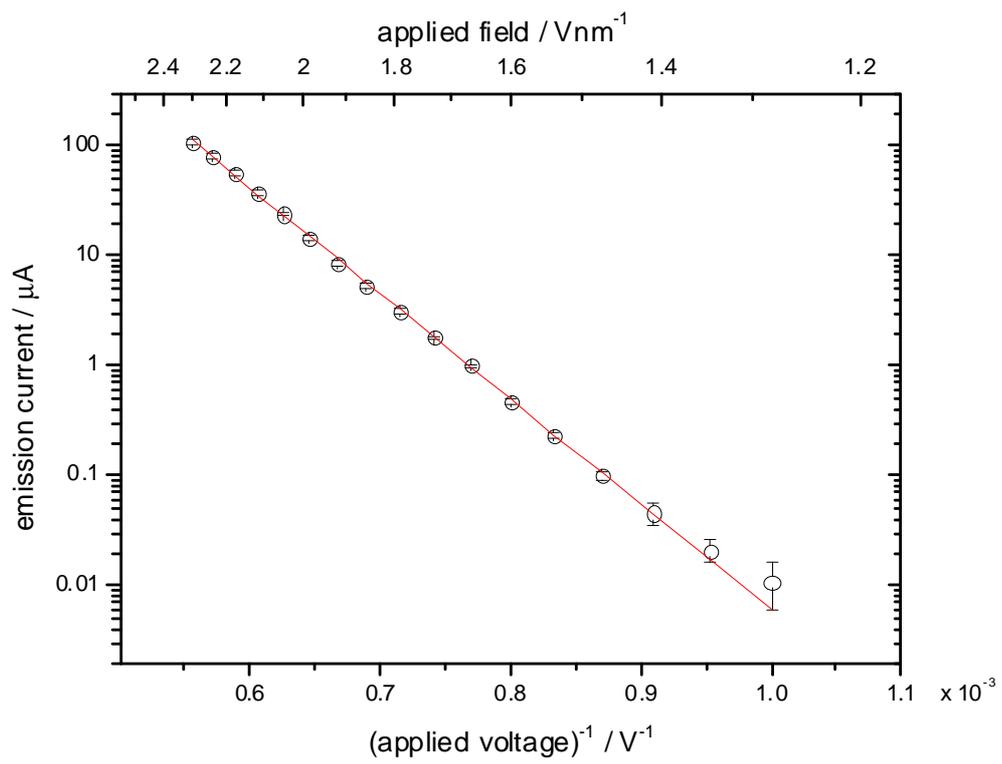